\documentclass[11pt,english]{article}

\usepackage{graphicx}

\usepackage{amsfonts}
\usepackage{url}
\usepackage[thmmarks]{ntheorem}
\newtheorem{theorem}{Theorem}[section]

\begin{document}
\title{Time of Philosophers, Time of Physicists, Time of Mathematicians}
\author{Fabien Besnard,\\ EPF}

\maketitle
\abstract{Is presentism  compatible with relativity ? This question has been much debated since the argument first proposed by Rietdijk and Putnam. The goal of this text is to study the implications of  relativity and quantum mechanics on presentism, possibilism, and eternalism. We put the emphasis on the implicit metaphysical preconceptions underlying each of these different approaches to the question of time. We show that there exists a unique version of presentism which is both non-trivial, in the sense that it does not reduce the present to a unique event, and compatible with special relativity and quantum mechanics: the one in which the present of an observer at a point is identified with the backward light cone of that point. However, this compatibility is achieved at the cost of a renouncement to the notion of an objective, observer-independent reality. We also argue that no non-trivial version of presentism survives in general relativity, except if some mechanism forbids the existence of closed timelike curves, in which case precisely one version of possibilism does survive. We remark that the above physical theories force the presentist/possibilist's view of reality to shrink and break up, whereas the eternalist, on the contrary, is forced to grant the status of reality to more and more entities. Finally, we identify mathematics as the ``deus ex machina" allowing the eternalist to unify his vision of reality into a coherent whole, and offer to him an  ``idealist deal": to accept a mathematical ontology in exchange for the assurance of surviving any physical theory.}

\section{Introduction}

Are the present events the only real ones ? Are the past, or even future events, also real ? This is one of the most debated question in the philosophy of time. We adopt here the terminology of Savitt (see \cite{savitt}): \emph{presentism} is the theory according to which only the present is real. In the opposite view, that of \emph{eternalism}, every event, be it present, past or future, exists in the same way. Finally, in the hybrid theory of \emph{possibilism}, the events which are in the past or the present, but not in the future, exist.

Let us review some of the arguments in favor of these different views, and some of their most obvious weaknesses.

Presentism is maybe the spontaneous philosophy of time. What is gone is gone, and the future is still open. Time passes, obviously. This theory would be the closest to common sense and intuition. It is also the most ontologically parsimonious. For a presentist, time is fundamentally different from space, and in order to support this claim, one could plead that it is possible to travel in space, but not in time. In particular, the passage of time is nothing like traveling forward in time. In a word, time is not a dimension.

The main argument of possibilism is the openness of the future, which renders it unreal. Symmetrically, what makes the past real is that we cannot change it. The passage of time is also part of this theory.

Since eternalism is immediately confronted with the seemingly solid arguments in favor of its rival theories, a word must be said on how it deals with them. First, from the eternalist point of view, the passage of time is not an objective  phenomenon. In Hermann Weyl's words (\cite{weyl}), ``Only to the gaze of my consciousness, crawling upwards along the life line of my  body, does a section of this world come to life as a fleeting image in space which continuously change in time". For an eternalist, the space travel vs time travel argument is a fallacy, for we never travel in space, but always in spacetime: we go from here-now to there-later. However, there appears to be nothing to answer to the argument of ontological parsimony. At first sight, the eternalist seems to offer far-fetched interpretations of obvious facts only for the sake of defending an idealistic view of reality,  completely remote from experience. Indeed, before the advent of special relativity, there were no convincing argument in favor of eternalism\footnote{Some even claim (e.g. \cite{bigelow}) that there were not a single eternalist before the late nineteenth century. Nevertheless,  the Eleatic philosophy, with its denial of change, could certainly be thought to be a form of eternalism.}. However, this theory provoked a dramatic change\footnote{According to Savitt (\cite{savitt}), eternalism is now the most popular theory of time among philosophers.} and has become the main weapon in the hands of eternalists. We will see why in section 3, by quickly reviewing the Rietdijk-Putnam argument.

But in order to understand the implications of relativity on presentism and possibilism, an operational definition of these points of view is first required. This is a notoriously difficult task. Indeed, McTaggart's argument (\cite{mactag}) might be a proof that presentism cannot be defined consistently only in terms of the ``A-series". We will escape the difficulty by formulating presentism inside a spacetime framework. Note that, from a presentist perspective, spacetime is to be understood only as a useful and purely mathematical tool. 

Once this definition is given, we will see that the Rietdijk-Putnam argument does not invalidate presentism outright, but forces it into incorporating the perhaps counter-intuitive idea that reality is observer-dependent. Therefore special relativity does not reduce presentism to nonsense, but rather clarifies this doctrine, by unravelling hidden assumptions. Some presentists might not be happy with this, but we will try to show that it is the only way out that is offered to them\footnote{In fact, there might be another one, which would be given by Stein's definition of the present as a single spacetime point (see \cite{stein}). We will only allude briefly to it in the text, since we believe it to be too remote from the general conceptions about the present to serve as a basis for a presentist theory. At least, it would not agree with the definition we give in the next section.}. 

Discussions about presentism are mostly centered on special relativity. We will show that general relativity is even more inhospitable to presentism, giving little hope that this doctrine could sensibly be held in this setting. Thus, we will argue that presentists should better fall back on possibilism. Though this point of view also has its own dfficulties with general relativity, we will argue that they are of a less severe kind, and could be cured if general relativity is supplemented with Hawking's chronology protection conjecture.

But there is more to this debate that the two theories of relativity: quantum mechanics must be taken into account. At first sight, this theory does not seem to have anything relevant to tell about spacetime, since it is tied to the old Newtonian views on that subject. Even if we pass to quantum field theory, it would seem that nothing is gained by comparison to the much simpler  theory of special relativity. However, the mere possibility of truly random events gives a lot of weight to the presentist/possibilist argument about the openness of the future. Could it be that our two most accurate physical theories to date each destroys a different philosophical theory of time, leaving the debate to be settled by the (hopefully) forthcoming theory of quantum gravity ?

We do not think so. Indeed, we will argue that the question under scrutiny is not a debate solely between philosophy and physics, but that mathematics also plays a major part. Indeed, it is clear that one cannot expect to answer the question about the reality of the past, the present and the future, if one does not give a sufficiently precise definition of reality beforehand. For instance, if one adopts a wholly mathematical ontology, that is, claiming that only mathematical objects are real, then eternalism simply follows. This might appear to be a big leap, but we will argue that it is a consequence of our analysis that eternalism leans towards this sort of ontology, whereas presentism and possibilism are naturally led to a purely empirical, positivistic, notion of reality. Thus, the aim of this paper might be summed up in the following may: to find the definition of reality which is coherent with the three main philosophical theories of time, given the metaphysical direction that the proponents of these theories must take to be able to stand their position in the context of modern physics.

\section{A Tentative Definition of Presentism and Possibilism}
The presentist theory of time appears to consist of two claims. The first, which is the most directly comparable with physical theories, is the following one: \emph{the world fundamentally has only three dimensions, each of which is spatial}.

\begin{figure}[hbtp]
\begin{center}
\includegraphics[scale=0.8]{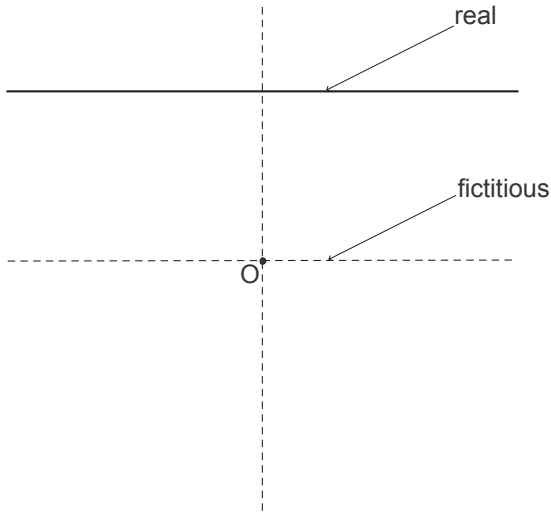}
\caption{Only the present is real.}\label{present}
\end{center}
\end{figure}

According to this view, spacetime representations of the world, such as those used in relativity, are nothing more than useful mathematical fictions. At any given moment, the whole reality\footnote{In all the article, when endorsing a presentist view, we do not distinguish between reality and the present. It means that we neglect other aspects reality might have, which are not relevant here.} is represented by a 3-dimensional subset of spacetime: the present. At this point, one has to be careful about not being too restrictive. In figure \ref{present}, the 3-dimensional subset representing reality is very special. There is no {\it a priori} reason to identify  the present with a spacelike hyperplane. Anything that gives meaning to the adjective ``3-dimensional''  has to be considered. Moreover, we set aside for the moment the issue of observer-dependence: this will be clarified later.

The second claim is that \emph{time passes}. Even though there is no general agreement about what this precisely means, it should be uncontroversial that it implies a clear-cut separation between the set of (real) present events, and the sets of (fictitious) past and future events. Moreover, each event is present exactly once. It means that the set of all presents (presents at different moments) realizes a partition of spacetime\footnote{At some point we will have to restrict ourselves to a partition of just a piece of spacetime.}. The movie depicting the evolution of the present would be an appropriate representation of the passage of time (in fact it would be even more appropriate to leave the present at the same position and let the rest of spacetime slide downwards).

Once the present has been defined, possibilism is most easily described as the belief that reality is the set of events that have been swept out by the present up to a given moment. In this way, each version of presentism gives rise to a corresponding version of possibilism. Conversely, a possibilist reality defines a presentist reality by taking the boundary. Accordingly, we will first mostly investigate presentism, viewing possibilism as a spin-off. However, we will eventually be forced to deviate slightly from the definitions that we have given in this section. At some point, we will have to consider a definition of reality that keeps the cumulative aspect of reality  which is inherent to possibilism, but is not associated with a companion presentist theory. Thus, we must remind that the definitions of this section are not definitive: they must be understood as a starting point.

\section{Presentism and Special Relativity}
\subsection{The Andromeda Paradox}
As we said in the introduction, presentism is notoriously difficult to combine with special relativity. Although it is now largely admitted that special relativity probably implies eternalism, or is at the very least a major challenge for the presentist theory, it seems that it took quite a long time to come to this conclusion since the birth of the theory in 1905\footnote{Some of the examples which follow are directly taken from \cite{petkov}.}. Some people almost immediately acknowledged the metaphysical implications of relativity for the philosophy of time, as is apparent in Minkowski's famous sentence\footnote{In \cite{mink}. I must confess that it is the only public commitment of Minkowski to the eternalist philosophy which I am aware of.} ``Henceforth space by itself, and time by itself, are doomed to fade away into mere shadows, and only a kind of union of the two will preserve and independant reality''. Weyl and G\" odel were also convinced. However, Einstein himself remained skeptical for quite a while. In an often quoted passage from his intellectual autobiography (\cite{carnap}, p. 37), Carnap reported Einstein's early 1950s worries about the Now. ``He explained that the experience of the Now means something special for man, something essentially different from the past and the future, but that this important difference does not and cannot occur within physics. That this experience cannot be grasped by science seemed to him a matter of painful but inevitable resignation''. On the other hand, in the 5th  appendix of the 15th edition of his classic popular book on relativity (\cite{eins}, p. 108), Einstein wrote ``Since there exists in this four dimensional structure no longer any sections which represent \emph{now} objectively, the concepts of happening and becoming are indeed not completely suspended, but yet complicated. It appears therefore more natural to think of physical reality as a four dimensional existence, instead of, as hitherto, the evolution of a three dimensional existence''.

In the purely philosophical field, the first argument aiming at a dismissal of presentism through special relativity is due to Rietdijk (\cite{riet}), in 1966, that is, more than sixty years after the advent of Einstein's theory. Strangely enough, it was almost immediately, and independently, put forward again by Putnam (\cite{putnam}). This argument is  also known as the ``Andromeda Paradox'' from its colorful presentation by Penrose (\cite{penrose}). Since Penrose's setting is a bit simpler, making use of only two observers instead of three in the original argument, this is the one we recall below.

Let ${\cal A}$ and ${\cal B}$ be two observers\footnote{Recall that in relativity, an observer is represented by his world-line. An inertial observer is then just the same thing as a timelike straight line in Minkowski space. In the sequel, when we say something like ``when ${\cal A}$ is at $O$'', we only mean that we consider a particular point $O$ on ${\cal A}$. However, one might want to imagine that in such circumstances we are really talking about someone, say Alice, who is experiencing a particular moment of her existence. This second interpretation is necessary when we take a presentist stance. As long as Alice's spatial extension can be ignored, it creates no difficulty, and one can freely pass from one interpretation to the other. Note also that in order to avoid any trouble with gender, from now on, the observer will be considered to be an asexual robot. Such an observer has the additional advantage of being possibly eternal, which is helpful for mathematical idealization.} passing each other in the street, at a relative speed of a few km/h. Their meeting defines a point in spacetime (that is, an event) that we shall call $O$. We assume that the observers are inertial\footnote{This is not really a loss of generality, since we can always use their local inertial frames at $O$ to carry the analysis, see subsection \ref{acc}.}. The observer ${\cal A}$ walks towards the Andromeda galaxy, and ${\cal B}$ goes in the opposite direction. In the simultaneity hyperplane of ${\cal A}$ one finds an event $M$, which is so defined: an Andromedean space fleet sets off to invade Earth. According to ${\cal B}$ when it is at $O$, $M$ has not yet taken place. It might even be that the decision to invade Earth has not already been made ! 

\begin{figure}[hbtp]
\begin{center}
\includegraphics[scale=0.8]{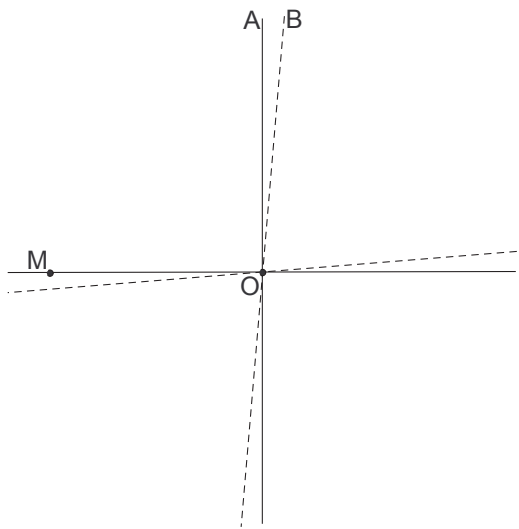}
\caption{The Andromeda paradox. The solid horizontal line represents the simultaneity hyperplane of ${\cal A}$ at $O$ (two spatial dimensions removed). The dotted lines are, respectively, the world-line of ${\cal B}$ and its simultaneity hyperplane at $O$.}
\end{center}
\end{figure}

What does the Andromeda Paradox really tells about presentism ? According to some (e.g. \cite{petkov}) it is more or less a \emph{reduction ad absurdum}: the Andromeda Paradox, as well as several other relativistic effects, are simply incompatible with a three-dimensional view of reality, hence with presentism. Savitt (\cite{savitt}) reviews the responses of some eminent presentists to this problem. We must confess that none of them appears to be very convincing to us. For instance, Arthur Prior\footnote{See \cite{prior}. See also Craig (\cite{craig}) and the refutation by Balashov and Janssen (\cite{bala}).} sees this paradox as an indication that relativity is an incomplete theory. It would only account for some physical impossibilities, or some appearances, not for the underlying reality\footnote{Bergson's own point of view was certainly of this kind.}. It can be thought of as withdrawal to a neo-Lorentzian position, which seems to be physically untenable.

However, if we use the definition of presentism that we have given above, the analysis of the paradox is quite straightforward. First of all, the formulation  implicitly defines the present of an inertial observer as the spacelike hyperplane which is orthogonal to its world-line. This is the usual definition of simultaneity in relativity, and it fits well with our requirement that reality should be a ``three-dimensional spatial thing''. There might be other choices, and indeed there are, as we will see, but for the moment let us stick with this definition of reality. Then the so-called paradox boils down to a simple syllogism: according to presentism, what is real for ${\cal A}$ when ${\cal A}$ is at $O$ is what is simultaneous with $O$.  But Einstein taught us that simultaneity is relative. Therefore reality is relative\footnote{The emphasis put on relativity in the very name of the theory is thus perfectly justified from the point of view of presentism, but not at all from that of eternalism !}. The Andromeda Paradox is an illustration of this fact, and also shows that two observers might be at the same point in spacetime and yet do not share the same reality. Thus, the presentist concept of reality must take the form of a relation between events, which depends on the observer (i.e. there is a different relation for each observer). Explicitly, if we write for short   

\begin{equation}
M{\cal R}_{\cal B}O\Longleftrightarrow \mbox{``}M\mbox{ is real for }{\cal B}\mbox{ when }{\cal B}\mbox{ is at }O\mbox{''}\label{relreal}
\end{equation}

then this relation can be immediately translated in the language of special relativity as

\begin{equation}
M{\cal R}_{\cal B}O\Longleftrightarrow  O\in{\cal B}\mbox{ and }M \mbox{ is simultaneous with }O\mbox{ with respect to }{\cal B}
\end{equation}

Let us clarify some properties of this family of relations. Simultaneity with respect to an observer is an equivalence relation, but clearly ${\cal R}_{\cal B}$ is not, since the two events do not play a symmetric role. Moreover, one could think of a possibility of exchanging the role of the events by switching observers, but this does not work either. For instance, let us define $G$ to be Galileo's death, and $N$ to be Newton's birth. Writing ${\cal G}$ for (the world-line of) Galileo and ${\cal N}$ for that of Newton, we can assume that (within some approximation)

$$N{\cal R}_{\cal G}G\mbox{ but not }G{\cal R}_{\cal N}N$$

That is, Newton's birth belongs to Galileo's reality when Galileo passes away, but Galileo's death is not real for Newton at the moment of his birth. This sort of phenomenon can occur if, and only if, Galileo and Newton have a nonzero relative speed.

Moreover, we learn from the Andromeda Paradox that even if $O$ belongs to the world-line of two observers ${\cal A}$ and ${\cal B}$, in general we have

$$M{\cal R}_{\cal A}O\not\Rightarrow M{\cal R}_{\cal B}O$$

All we have seen so far is that the presentist is forced to adopt a definition of reality which depends on the observer, and more precisely on the movement of the observer. Moreover, the relation of reality with respect to an observer has peculiar and maybe unexpected properties. But this is by no means a reduction of presentism to pure nonsense. However it should be noted that many authors whose works argue for the incompatibility between presentism and relativity (\cite{putnam}, \cite{petkov}, \cite{savitt2}, \cite{saunders}) explicitly reject the idea of an observer-dependent reality. Indeed, it is clear that one cannot hold the three following theses for true at the same time:

\begin{enumerate}
\item\label{t1} There is an objective, observer-independent reality.
\item\label{t2} Philosophy of time has to be compatible with special relativity.
\item\label{t3} Presentism is the correct theory of time.
\end{enumerate}

We think that it is worth emphasizing that the arguments against presentism put forward by Putnam, Savitt, Saunders, and many others, all rely on a commitment to \ref{t1} and \ref{t2}. On the other hand, it seems to us that most philosophers who wish to stick to presentism tend to eliminate \ref{t2}, as we noted above. We think that it is a weak line of defense, and that keeping \ref{t2} while getting rid of \ref{t1} would be much more reasonable, opening the road for a theory of ``relative presentism''.

However, even if one opts for such a relative presentism, the peculiar properties of reality that we have underlined above might seem undesirable. Since the analysis, as we have remarked, is tied to a particular definition of simultaneity in special relativity, this raises the following question: can one change that definition in order to obtain a better behaved notion of reality ?

\subsection{Simultaneity Conventions in Special Relativity}
\begin{figure}[hbtp]
\begin{center}
\includegraphics[scale=0.8]{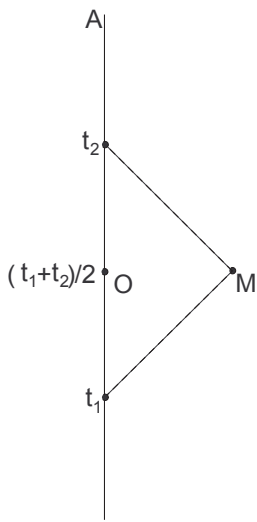}
\end{center}
\caption{Poincar\'e-Einstein convention of simultaneity}\label{pesim}
\end{figure}

The usual way to define simultaneity in Minkowski spacetime is called \emph{standard synchrony} or Poincar\'e-Einstein simultaneity. It can be described as in figure \ref{pesim}: an observer ${\cal A}$ sends a light signal at time $t_1$, as indicated by its wristwatch\footnote{The procedure described here seems to require the use of some clock carried by ${\cal A}$, but in fact it is sufficient that ${\cal A}$ knows how to define the midpoint of a segment on its world-line, and this can be achieved without a clock thanks to the causal structure of Minkowski spacetime. This is part of Malament's theorem, to be discussed below.}. The signal is reflected back to ${\cal A}$ by a mirror at the spacetime point $M$, and then received by ${\cal A}$ at time $t_2$. The event $M$ is then assigned the time coordinate $t={1\over 2}(t_1+t_2)$. Let us call $O$ the event with time coordinate $t$ on the world-line of ${\cal A}$ (this event is physically defined by the position of the hand of ${\cal A}'s$ watch). The procedure  defines the events $M$ and $O$ to be simultaneous.

Natural as it may seem, this definition was considered a pure convention since Einstein's days. This is the conventionalist view on simultaneity, which has been fully developped in \cite{reich}. We will not enter into the details of this debate, in particular the alternative $\epsilon$-definitions of simultaneity introduced by Reichenbach (we refer the interested reader to \cite{janis} for an overview). Let us only remark that this issue is of major importance for the presentist/possibilist philosopher: even if one is ready to discard the postulate that reality is independent of the observer, we think it is safe to assume that, at the very least, reality should not be a matter of convention. As a result, were the conventionalist thesis true, it would suffice to invalidate the presentist/possibilist theory. This is why we will put the emphasis on a result obtnained by Malament \cite{malam}, which, contrarily to most expectation, could be interpreted as saying that the conventionalist thesis was wrong ! Malament  proved that, if we agree that a definition of simultaneity should \emph{not} be called conventional if it is entirely definable in terms of the causal structure of Minkowski spacetime, then Poincar\'e-Einstein simultaneity is essentially unique. In order to be more precise, let us introduce some notations.

First, given two events $M$ and $N$, we say that they are causally connected, and we write $M\kappa N$, if, and only if, they can be joined by a curve the tangent vector of which is always lightlike or timelike\footnote{Of course in Minkowski spacetime this curve can always chosen to be a straight line, but as it is written, the definition generalizes to curved spacetimes with no modification.}. This is equivalent to saying that each event belongs to the full light cone of the other. We also need to define a causal automorphism: this is a bijection $f$ of Minkowski spacetime into itself, such that $M\kappa N$ iff $f(M)\kappa f(N)$. We can now state Malament's theorem:

\begin{theorem}\label{malamtheorem} {\bf (Malament, \cite{malam})}

Given the world-line ${\cal A}$ of an inertial observer, the only relation which is:

\begin{enumerate}
\item an equivalence relation,
\item non trivial (all spacetime points are not equivalent, and at least a point on ${\cal A}$ is equivalent to a point not on ${\cal A}$),
\item invariant by all causal automorphisms stabilizing ${\cal A}$,
\end{enumerate}

is Poincar\'e-Einstein simultaneity relative to ${\cal A}$.
\end{theorem}

The last requirement is a necessary condition for a relation to be definable from $\kappa$ and ${\cal A}$ alone. However, Malament observed that conversely, it is easy to see that Poincar\'e-Einstein simultaneity is indeed definable from these data. 

This theorem is good news for the presentist/possibilist side. However, it has been extensively criticized and even allegedly refuted by some (see \cite{sarkstach}, \cite{benyami}, \cite{giul}, and \cite{bes}). The critics mostly concentrated on the use of the symmetric relation $\kappa$ instead of the causal order relation $M\preceq N$, which is defined by $M\preceq N$ if, and only if, there exists a curve joining $M$ to $N$ the tangent vector of which is always timelike and future directed (this requires a time orientation). Of course $\preceq$ is a stronger piece of data than $\kappa$, so the question is: how many additional structures do we require to be preserved ? We will leave aside the part of this question which is purely internal to relativity theory. We certainly need to enter this debate, but only as far as the presentist/possibilist theories of time are concerned. From this stance, it will suffice to observe that the passage of time, which is part of both presentism and possibilism, requires a distinction between past and future at every event on the observer's world-line. Thus we think it is  legitimate to add the following structure: to each point $x$ on ${\cal A}$'s world-line are associated two subsets $\uparrow x$ and $\downarrow x$, respectively called the future and the past of ${\cal A}$ at $x$, and such that:

\begin{itemize}
\item $\{x\}$, $\uparrow x$, and $\downarrow x$ form a partition of ${\cal A}$,
\item if $y\in\uparrow x$ and $z\in\uparrow y$ then $z\in \uparrow x$.
\end{itemize}

Up to this point, this data is equivalent to a total order relation $<$ on ${\cal A}$ defined by $x<y\Leftrightarrow y\in\uparrow x$. However, this order relation is completely arbitrary and in particular not necessarily compatible with the natural topology of ${\cal A}$ (in other words $\uparrow x$ can be any subset of ${\cal A}$). Thus we think it is justified to restrict to the cases where $\uparrow x$ is a half-line of ${\cal A}$. It then easy to see that there are only two possibilities, each one corresponding to an orientation of ${\cal A}$. We will call the corresponding order relations \emph{natural orderings}. It is then possible to show the following generalization of Malament's theorem (the proof of this theorem, and of all the preceding claims as well, can be found in \cite{bes}):

\begin{theorem}\label{letheorem}
Let ${\cal A}$ be an inertial observer and $<$ a natural ordering on ${\cal A}$. If $R$ is a relation on Minkowski spacetime such that $R$ is
\begin{enumerate}
\item an equivalence relation,
\item not trivial,
\item definable from ${\cal A}$, $\kappa$, and $<$,
\end{enumerate}
then  either $R$ is   the Poincar\'e-Einstein simultaneity relation with respect to ${\cal A}$, or it is defined by the partition of Minkowski spacetime into half cones with their apices on ${\cal A}$. Conversely, all these relations are definable from the given data.
\end{theorem}

The Poincar\'e-Einstein relation will be abbreviated PE in what follows. The other kinds of relations will be called ``conic relations".  Among them we distinguish the relation of ``observed simultaneity" (abbreviated OS), which is defined by the backward light cones\footnote{The notion of backward light cone is easily defined using $<$ and the causality relation: $y$ is in the backward light cone of $x\in{\cal A}$ iff $y$ is in the light cone of $x$ and the light cone of $y$ intersects ${\cal A}$ at $x'<x$.}. The relation defined by the forward light cones is the dual of OS, which means that it is OS relative to the dual ordering $>$.

Apart from OS and its dual, there is an infinity of other conic relations, for which the simultaneity classes are half cones generated by a timelike or a spacelike half line. These other conic relations depend both on the position and the movement of the observer, and consequently cumulate the drawbacks of PE and OS, without any advantage. We will not discuss them in this paper because, on the one hand, we see no reason why one would like to use such unnatural definitions, and, on the other hand, the arguments we will give in the sequel, coming from general relativity and quantum theory, in order to discard PE presentism/possibilism, would also apply to these hybrid versions. 

As far as relativity theory only is concerned, there is no reason to prefer OS to its dual. However, we will assume in the rest of the paper that some time arrow (for instance the electromagnetic time arrow, or the time arrow defined by the formation of memories in the observer's brain) is given. This additional information, that we certainly cannot ignore while discussing the philosophical theories of time, forces us to discard the dual of OS as a possible definition of simultaneity. Thus, we will not refer to the dual of OS anymore in the sequel. It would nevertheless be easy for the reader who so wishes to erase every reference to the time arrow in what follows by ``dualizing" every observation which applies to OS.  

Thus, we end up with two reasonable definitions of reality for the presentist: the one coming from PE, and the one coming from OS, and for the rest of this section we will compare their merits.  Let us call the set of points simultaneous with $O$ for ${\cal A}$ the reality of ${\cal A}$ at $O$. Realities according to PE and OS have very different properties.

\begin{itemize}
\item According to PE, if ${\cal A}$ and ${\cal B}$ have parallel world-lines, then the set of all realities of ${\cal A}$ coincides with the set of realities of ${\cal B}$. Else, no reality of ${\cal A}$ ever coincides with a reality of  ${\cal B}$, even at the intersection point of the world-lines of ${\cal A}$ and ${\cal B}$, in case they intersect (Andromeda Paradox).
\item According to OS, the reality of ${\cal A}$ at $M$ and the one of ${\cal B}$ at $N$ coincide if, and only if, $M=N$.
\end{itemize}

To sum up, reality according to PE depends on the movement of the observer, whereas according to OS it depends on the position of the observer in spacetime. Note that, by definition, the reality of an observer always depends on its position \emph{along} its world-line. With PE it also depends on the direction of the world-line, whereas with OS it does not depend on the world-line at all. As Sarkal and Stachel remarked, Einstein did consider the OS definition of simultaneity, but discarded it on the ground that it depended on the position of the observer in space. However, Einstein's goal was not to find a correct formulation of presentism. With this objective in mind, it might appear more natural that the dependence of reality on the observer goes through the observer's position rather than through its movement. We can also note that with PE, the presentist reality is reconstructed: using the notations of the beginning of this section, ${\cal A}$ knows at time $t_2>t$ that $M$ \emph{was} real when its watch read $t$. On the contrary, using OS, one gets an immediate, almost obvious, presentist definition of reality: \emph{what is real for me now is exactly what I can see now}\footnote{This is to be taken {\it cum grano salis}: it is not meant here that objects are not real when we are not looking at them. In fact neither vision nor even the propagation of light play any particular role in this definition. The only things that count are the causal structure of spacetime and a given time orientation.}.

However, it important to realize that the hypotheses of theorems \ref{malamtheorem} and  \ref{letheorem}) which permit us to single out PE and OS as the only reasonable definitions of reality from a presentist/possibilist stance, do not allow us to use a clock. This does not matter for OS, as we have remarked above, but it would pose a serious problem for a PE presentist/possibilist since it is impossible to give an operational meaning to this definition of reality. Indeed, the geometrical definition of PE  uses only the causal connectibility relation and the observer's world-line, but how is the observer supposed to know  that $M\kappa N$ is true when $M$ and $N$ do not both belong to a light cone centered on its world-line ? To acquire such information angular and duration measurements are needed. Indeed, it is possible to show that if we restrict the hypotheses of  theorem \ref{letheorem} by replacing $\kappa$ with the ``local'' light cone structure, that is the set of half light cones with their apices on ${\cal A}$, then OS and its dual are the only relations remaining (see \cite{bes} for a proof of this maybe obvious statement). However, if we equip the observer with a clock and a laser pointer, it can then cook up a completely arbitrary simultaneity relation.

On the contrary, the OS definition does not require any tool. It is the only relation which only requires the observer to observe ! We see that, when we take an operational point of view, we are very far from singling out PE as the unique simultaneity relation, thus the only natural presentist definition of reality. Quite the contrary: either we equip the observer with enough tools to construct any simultaneity relation it likes, or we restrict to the least possible data compatible with the definition of presentism/possibilism, and OS comes up as unique. In the first case, conventionalism is true and presentism/possibilism are both dead. In the second case, PE presentism/possibilism are dead. However, we wish to push the analysis as far as we can, and in the next sections we will still consider PE presentism/possibilism (if only to put more nails in its coffin). After all, this definition is still the most popular and is deeply routed in our habits, so we need really good reasons to dump it.

The OS definition of the present, for which we have argued above, has been considered before. Putnam immediately rejects it (\cite{putnam}) because it would violate the principle that no observer is privileged in relativity. However, we think that any observer is justified to see itself as privileged\footnote{It appears clearly in the hypotheses of the theorems proved in \cite{malam}, \cite{sarkstach}, \cite{giul}, for instance.} when it comes to defining its own reality. What Putnam really rejects is that reality could be observer-dependent. The analysis of OS by Saunders (\cite{saunders}) comes to the same conclusion. Savitt (\cite{savitt2}) has a different point. For him, any set of events used to define the present should be \emph{achronal}. It means that it should not contain any couple of events $(A;B)$ such that $A$ happens before $B$ according to all inertial observers. But this definition is equivalent to saying that the present should not contain any couple $(A;B)$ such that $A$ is in the full backward cone of $B$. This is more or less a way to discard OS by definition. Furthermore, the definition of ``achronal'' uses the chronology defined by inertial coordinates: this is the PE convention in disguise, it is not adapted to OS. It is perfectly possible to use coordinates which are adapted to OS, for instance light cone coordinates, which are well known in general relativity.  


Anyway, OS presentism is a consistent doctrine, though not very popular. It  has been endorsed by some philosophers, and the reader might like to consult \cite{godsmith} or \cite{benyami} to learn more about it.

\begin{figure}[hbtp]
\begin{center}
\parbox{7cm}{\includegraphics[scale=0.65]{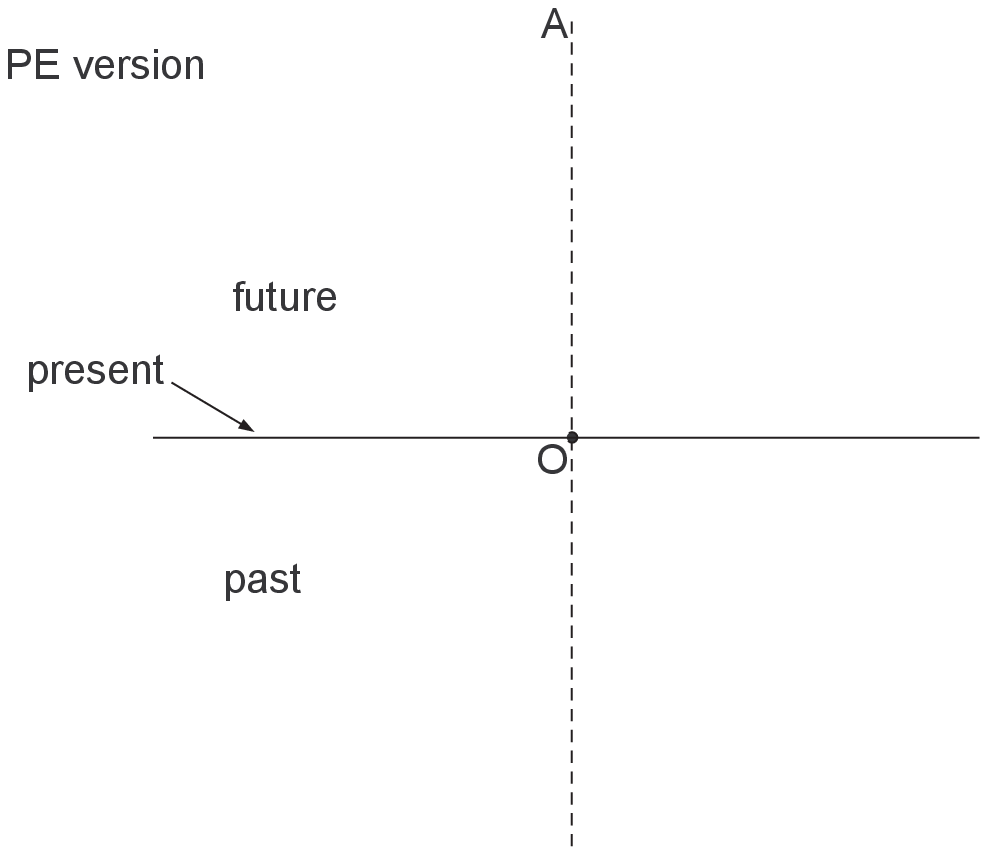}}\parbox{9cm}{\includegraphics[scale=0.65]{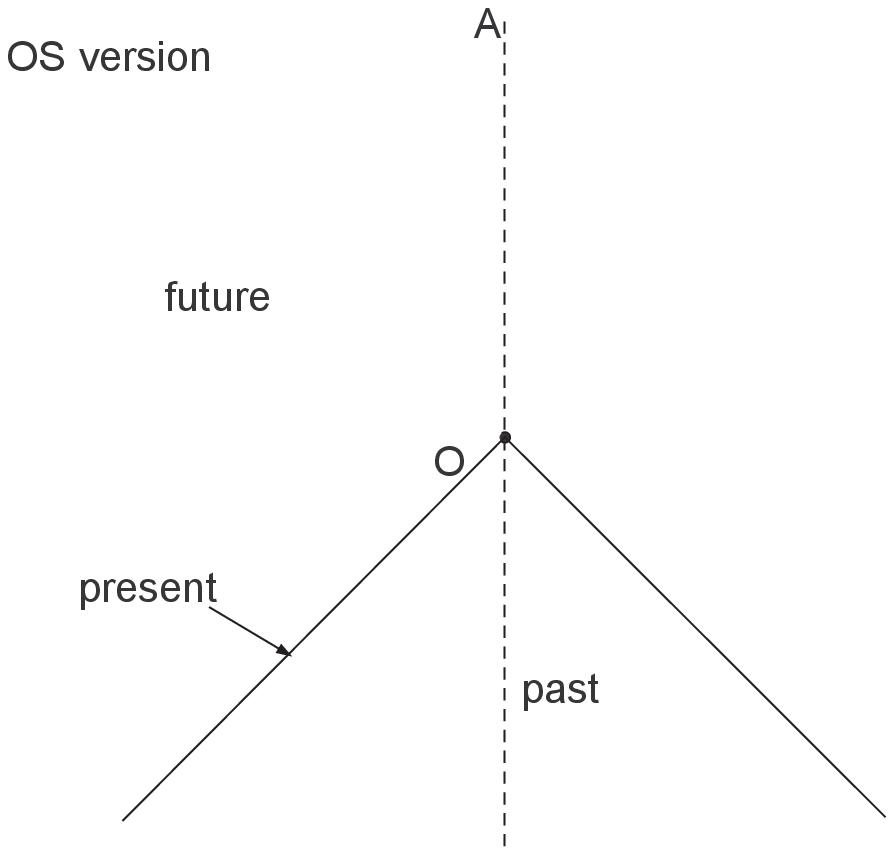}}
\caption{Representations of the past, present and future at $O$ of an observer ${\cal A}$, according to PE and OS, respectively.}\label{passfut}
\end{center}
\end{figure}

To close this section we have to say a word about the past (which is needed to define possibilism) and the future. These notions follow immediately once we know what the present is. There are two ways of defining them. First, we can say that the present is the border between past and future, and this determines the latter up to the choice of a time arrow. The second way is to use the ordering given by proper time along the world-line of the observer, and propagate it thanks to the simultaneity relation. More precisely, an event is declared to be in the past if it is simultaneous with an event in the past on the world-line ${\cal A}$. The future is defined similarly. The two procedures amount to the same, and the result is summarized in figure \ref{passfut}. Remark that with the OS definition, the future includes what is generally called the ``elsewhere'' in relativity. We see that in OS possibilism, reality is defined in terms of causality. An event $M$ is real for an observer at $O$ if something at $M$ can influence this observer when it is at $O$.

\subsection{Accelerated Observers}\label{acc}
For the moment we have only dealt with inertial observers. But of course no observer can be perfectly inertial. It is therefore crucial to generalize what we have done to a realistic situation. Moreover, even if the debate on the compatibility  between presentism and special relativity mainly focused on inertial observers, it is worth noticing that accelerated observers can be treated without problem in this theory. In fact, the theory was even built with this aim ! Thus, the study would be rather incomplete if we did not cover this case.

Fortunately, both PE and OS can be straightforwardly generalized. For OS this is obvious, since this version does not depend at all on the movement of the observer. To define the present at $O$ of a general observer ${\cal A}$   (i.e. ${\cal A}$ is any timelike curve) with PE, we simply use the definition of the inertial observer ${\cal A}'$ which passes through $O$ with the same speed as ${\cal A}$ at this point (i.e. we replace ${\cal A}$ by the timelike straight line ${\cal A}'$ which is tangent to ${\cal A}$ at $O$). However, we immediately run into trouble, since the simultaneity hyperplanes will intersect.

\begin{figure}[hbtp]
\begin{center}
\includegraphics{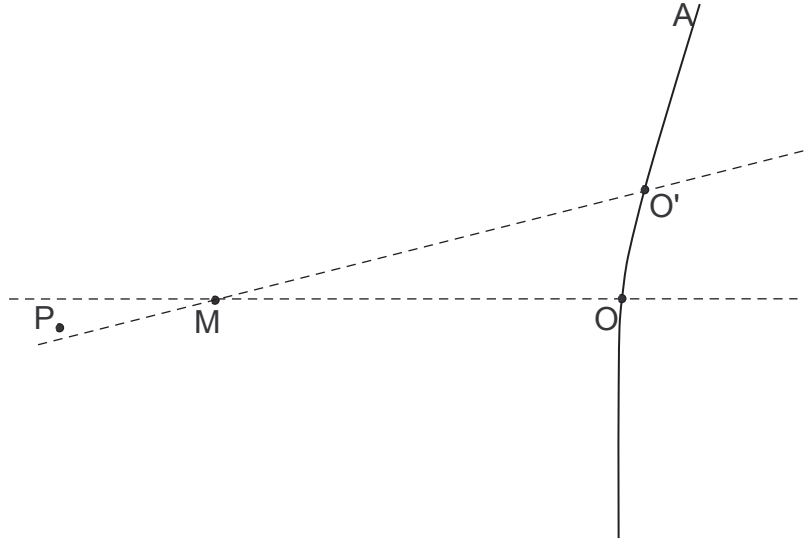}
\caption{An accelerated observer experiencing troubles with PE simultaneity.}\label{accel}\end{center}
\end{figure}

In figure \ref{accel}, we have drawn the world-line of an observer ${\cal A}$ whose movement is first inertial, then accelerated between $O$ and $O'$, then inertial again. We see that the event $M$ is present for ${\cal A}$ both at $O$ and at $O'$. Worse still, there are events ($P$ is an example) that are in the past for ${\cal A}$ at some point and in the future at a later point ! These features conflict with the very intuition behind   presentism. If ``what is gone is gone'' an event could not possibly be present twice. From a possibilist stance, it is unthinkable that an event might slip from the set of fixed and real events into the open future. From another point of view, if we insist that simultaneity must be an equivalence relation defined on all spacetime, then we are forced to declare that all events are simultaneous, thus ruining presentism and possibilism alike. The way out of this problem for the PE presentist/possibilist is to acknowledge that the present is not only a relative but also a \emph{local} notion, exactly like the coordinates. Recall that a PE presentist has to reconstruct his reality at $O$. Should this reality be spatially infinite, this reconstruction process would take an infinite time to complete. Now whenever  the information about the intersection of two of his simultaneity hyperplanes (an event like $M$) reaches him, he must set a spatial bound to his reality at $O$.  

\begin{figure}[hbtp]
\begin{center}
\includegraphics[scale=0.8]{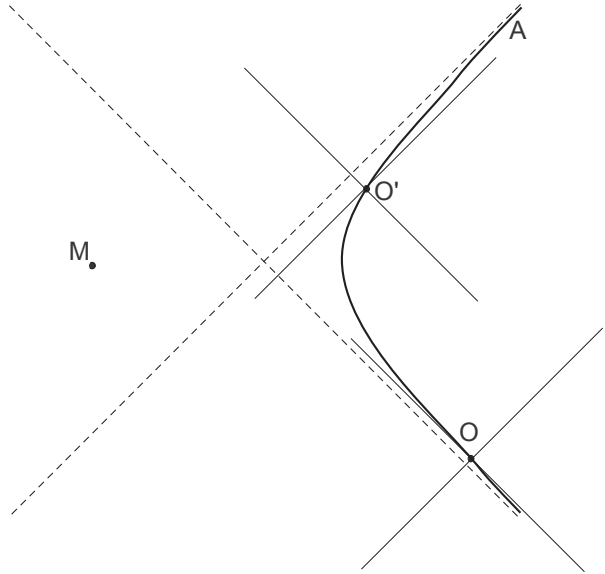}
\caption{No signal can reach a point on ${\cal A}$ from $M$. No signal sent from a point on ${\cal A}$ can ever reach $M$.}\label{unifaccel}\end{center}
\end{figure}

The OS presentist and possibilist theories are not, in special relativity, bothered by the intersection of the backward light cones of an observer, since this cannot happen, provided that the speed of the observer never exceeds $c$. However, in this version too one has to accept that, for some observers, reality can be local. By ``local'' here we mean that the set of realities of this observer does not cover all of Minkowski spacetime. Consider for instance a uniformly accelerated observer ${\cal A}$. Then there are events, like $M$ on figure \ref{unifaccel} that are never in the present, past or future of ${\cal A}$, as defined with the OS convention.

\section{Presentism, Possibilism, and Eternalism in General Relativity}

First, what are the differences between the two theories of relativity ? For some, the special theory is only the study of particular solution of the general one, so that the distinction between the two is more of a historical than a conceptual nature. However, we think that there is an important conceptual difference between the two. Inside the special theory, the metric tensor $\eta$ is not seen as a solution of some dynamical equations. It is a fixed background structure. Thus, one is allowed to use this structure to distinguish a privileged class of observers (the inertial observers) and a particular type of coordinates (those in which the metric takes its usual form).

Thus, the main new feature of general relativity is that the metric is not a background structure anymore: any metric satisfying Einstein's equations is allowed, and it generally has non-vanishing curvature. Another aspect that may play a role is that general relativity is a completely local theory which does not fix the global topology of spacetime.

These two new aspects entail that coordinates in general relativity are both arbitrary and local. There is now no way to escape the conclusion that simultaneity, in the spirit of PE, is conventional. Indeed, any foliation of (a piece of) spacetime by spacelike hypersurfaces furnishes a perfectly acceptable notion of simultaneity. In Minkowski spacetime we had the privileged inertial observers at our disposal. If we took the world-line ${\cal A}$ of one of them to define simultaneity at a point $O$, this singled out all parallel world-lines. The spacelike hyperplane of PE simultaneity of ${\cal A}$ at $O$ was then defined as the unique spacelike hypersurface $\eta$-orthogonal to all these world-lines and passing through $O$. This existence and uniqueness result is what had allowed us to pass from the local to the global in Minkowski spacetime. It relied on the existence of a partition of spacetime into straight lines parallel to ${\cal A}$. But there is no way of defining unambiguously distant parallelism in general relativity, precisely because the curvature does not vanish. Some solutions of Einstein's equations do single out a class of privileged observers, but in general the data of the world-line ${\cal A}$ of such an observer, and a point $O$ on it,  do not suffice to canonically generate a spacelike hypersurface. Indeed, there are an infinite number of spacelike hypersurfaces intersecting ${\cal A}$ orthogonally at $O$, and each one of them can be used to define Gaussian (or synchronous) coordinates which are a possible generalization of Poincar\'e-Einstein simultaneity. This renders PE presentism/possibilism untenable.

The OS version of these theories is much better behaved. 
The motto ``what is real is what I see'' is of course still physically well defined and keeps its meaning in general relativity. The reality for OS presentism would be what is sometimes called the \emph{past horismos} (\cite{beem}, p. 289). The past\footnote{The word ``past'' here is misleading, since in OS presentism/possibilism, the past horismos is precisely the present.} horismos of $O$ is the set of lightlike geodesics which end at $O$. We will continue to call this object ``the backward light cone of $O$'', even though, strictly speaking, the light cone lives in the tangent space. Since we will never use the tangent space explicitly, this little misnomer should cause no trouble. The OS possibilist version of reality for an observer at $O$ would be the causal past of $O$. However, peculiarities arise, as we shall  see, because light cones can now intersect. 

Indeed, that a single event may be perceived several times by an observer is  rather common in the Universe. Time delays between the multiple images obtained by gravitational lensing is a well-known phenomenon, which can even be used to determine the Hubble constant\footnote{Typing ``Time Delay Lensing'' returns dozens of relevant papers in Google Scholar. One can try \cite{schechter} for an overview of the subject.}. If we imagine a very special distribution of masses, it would even be possible for an observer to see its own birth. If the Universe has a cylindrical topology in the spatial directions, and this is a real possibility, it is not even needed to imagine a complicated distribution of matter: an observer shall see its own birth periodically (provided it lives long enough). This of course clashes with the presentist view of reality. If an event can be at the same time past and present, i.e. past and real, it is legitimate to declare democratically all past events real, that is adopting possibilism. It would seem rather arbitrary to go around this problem by defining an event to be real only the first time you see it. As a matter of fact, in some cases there might not even  be such a first time. It should be possible to limit the extension of reality of an observer to a neighborhood such that backward light cones do not intersect\footnote{It is proved in \cite{sachs} that so-called ``convex normal neighborhood'' have this property.}. However, the clear-cut physical interpretation of OS presentism would be broken, and it too would seem rather arbitrary. 

But is OS possibilism free of problems ? One might argue that, taking the same example as above, at the time of my birth, this event can be both present and future for me. This would make real an event from the future, thus ruining possibilism. However, note that in case the light emanating from my birth is redirected on me by some particular distribution of matter,  I cannot know at the time of my birth that such a thing will happen. In fact, I can be informed of this fact only when I meet that light again\footnote{To calculate the path taken by the returning light, I would have to know the Christoffel symbols in the direction of the tangent vectors to this path at every point of this path, and I can't have this information before the light returns to me.}. But more fundamentally, the possibilist vision of reality is that of a growing reality (a growing universe, some say) which should not be troubled by the ``multiple birth paradox''. The possibilist reality of an observer at $O$ is  just the causal past of $O$. The birth of the observer clearly belongs to this set, it is not a problem if the observer is informed several times of this fact.

However, there is another threat coming from certain solutions of general relativity, first discovered by G\" odel, in which there exist closed timelike curves (CTC). On such a curve, any point is both in the past and in the future of any other. Since no distinction between future and past has any meaning, there can be no consistent possibilist theory if CTC's are allowed. The general theory of relativity certainly allows them, thus we can safely conclude that this theory is only compatible with eternalism. However, it is not clear whether Einstein's theory should be taken seriously on this point. For some\footnote{See for instance Hawking's chronology protection conjecture (\cite{hawking}).}, this is a pathology that ought to be cured in quantum gravity. Thus, if it turns out to be true, general relativity has to be supplemented with an extra assumption forbidding CTC's and OS possibilism would be consistent with this theory, that is general relativity with unphysical solutions removed.
 
\begin{figure}
\begin{center}
\includegraphics{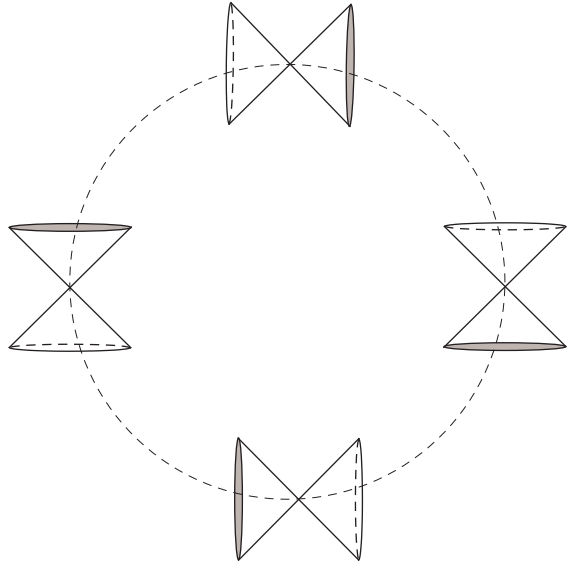}
\end{center}
\caption{A closed timelike curve and some light cones with their apex on it.}
\end{figure}

Finally, the question of locality acquires a new aspect in general relativity. In the special theory, some observers, like the uniformly accelerated ones, have event horizons, but in the general theory there exist absolute horizons. Consider for example the simple case of the Schwarzschild solution. An observer not going inside the event horizon will never receive any signal from behind it. An observer who crosses the horizon will still be reached by signals coming from the exterior, however it will hit the singularity in finite proper time. After that, the theory does not tell us what happens. But one can avoid this complication by considering a solution containing more than one black hole (there are many evidences that our spacetime actually contains a lot of them). In this case it is clear that no observer can have access to the totality of spacetime via its backward light cones. The ontological status of the regions of spacetime that a given observer is never informed about is problematic. From a possibilist stance, the events in such regions are never in the present/past of the observer, thus they can never be considered as real. 

Should an eternalist accept the idea that the whole of spacetime is real ? As far as Minkowski spacetime is concerned, this is the very definition of this doctrine. However, accepting the reality of events which are behind an absolute horizon requires a new inductive leap, since no direct evidence that such a region exists can ever be given. This leap is, however, very natural from a mathematical perspective, and, on the contrary, a punctured spacetime would seem less credible. In fact, all specialists in general relativity make this leap, at least as far as the theory is concerned. Accepting the idea that a more general spacetime that Minkowski's unrealistic one exists should then cause no extra trouble. Another point is that, even if  a single observer cannot acquire informations about events hidden behind the many black hole horizons of our universe, a particular observer can get to know what is lurking behind a particular horizon by simply diving into it. Note that for an eternalist the future and the past, being as real as the present, are being observed by future and past observers, respectively. The eternalist creed is that there is an observer-independent reality taking all these partial observations into account. The reasoning would be the same for black holes.  

Thus, we see that the general theory of relativity seems to tolerate only two points of view: eternalism, and OS possibilism, the latter only if some mechanism forbids CTC's. We can also notice two opposite trends emerging: a ``positivist'' trend for the possibilist, and an idealistic trend for the eternalist. The possibilist view of reality narrows down to a strict empiricism, whereas the eternalist is more and more inclined to believe in an exact   parallelism between reality and mathematical physics. This opposition will keep growing in what follows.

\section{Presentism, Possibilism,  and Eternalism in Quantum Mechanics}

One of the main arguments in favor of both presentism and possibilism is that of an open future. If we were to analyze where this intuition, shared by many, comes from, we would certainly find that it has two origins. The first is the belief in free will. It is not the place here to embark in a discussion on the relevance of this belief. However, it is worth noticing that the so-called paradoxes of time travel, like the grandfather paradox, fade away if we agree that our actions are constrained by the consistency of histories (see \cite{novikov}). Anyway, we think it is justified to say that there is no room in the scientific theories about the universe for a concept such as free will. Therefore, since this is an essay about the compatibility between the philosophical theories of time and modern physics, we will not consider the issue of free will any further. There remains the second origin to the intuition that the future is ``not yet fixed'', and that is chance. If truly random events exist, the idea that the future is open gains force, and it renders eternalism less credible. Since in principle no truly random events take place in the world of classical mechanics, we now turn our attention to quantum mechanics.

Although randomness is the only characteristic of quantum mechanics that we will consider, we do not mean to imply that other features, like entanglement for instance, have no impact on the question of time, or that quantum field theory, the successful mixture of special relativity and quantum mechanics, do not deserve special attention. Quite the opposite. However, we think that these questions are worth a study of their own, and we wish to keep this article within reasonable bounds.

In order to investigate the implications of randomness on the question of time, let us imagine a quantum heads-or-tails machine. It is easy to build: we take a spin one-half particle, an electron for instance, prepared in a pure state $|+,x\rangle$. It means that if we measure the spin of that particle along the $x$ axis, it will yield $+\hbar/2$ with complete certainty. However, if we measure it along the orthogonal $z$ axis, quantum mechanics (and experimental evidence) tells us that there are equal probabilities to find $+\hbar/2$ or $-\hbar/2$. After the measurement, the particle will be in the state $|+,z\rangle$ or $|-,z\rangle$, and in both cases, if we measure the spin along the $x$ axis again, we will find $+\hbar/2$ with probability $1/2$, or $-\hbar/2$ with probability $1/2$. Thus, a sequence of measurements along the $z$ and $x$ axis, in alternation, will yield a random sequence of $\pm\hbar/2$ results. We do not bother for the moment with any issue concerning the interpretation of quantum mechanics. We can even put the system into a box and forget about quantum mechanics: all that matters is that we now have a quantum coin to play heads-or-tails. The only assumption we make is that the results are truly random, and by this we mean that the randomness cannot be analyzed away by using some hidden variables.

\begin{figure}[hbtp]
\begin{center}
\includegraphics[scale=0.7]{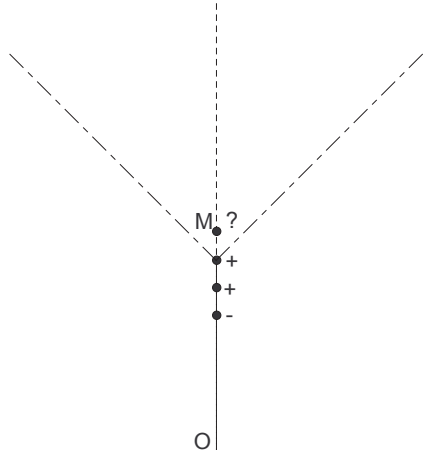}
\caption{A game of quantum heads-or-tails.}\label{qht}
\end{center}
\end{figure}

We then consider an observer ${\cal O}$ who plays with the quantum coin at moments separated by a constant amount $\Delta\tau$ of proper time. At the instant $t$, as measured by ${\cal O}$'s clock, the quantum coin is tossed, and the result is instantaneously recorded by ${\cal O}$ (it is the third event from the bottom on figure \ref{qht}, marked with a $+$). Since the next result (event $M$) is completely undetermined, we can argue that it belongs to an open future. It may be that the result of the toss will completely change the course of history (${\cal O}$ might be an Andromedian using the coin to determine if he will launch an attack against the Earth). Thus, the full forward cone of $M$ is equally open. Now if $\Delta\tau\rightarrow 0$ we see that  all the interior of the full forward cone of the event marked by a $+$ can be rightly be called an open future.

The metaphysical postulate which lies behind the idea that the ``future does not exist because it is still open'' is the following: something that cannot be determined, by any means, does not exist. This postulate of course needs to be relativized with respect to an observer. We call it the \emph{positivistic postulate}. We state it here more precisely and discuss its meaning below.

{\bf Positivistic Postulate (weak form):} {\it Something which cannot be determined by an observer ${\cal O}$ at $M$, even in principle, is not real  for ${\cal O}$ at $M$.}

The strong form would be ``Something which cannot be observed by ${\cal O}$ at $M$ is not real for ${\cal O}$ at $M$'' . What we will show is that, given the existence of a quantum coin, the weak form implies the strong form (which in turn obviously implies OS presentism or possibilism). Before doing so, it is important to realize that, conversely, in the absence of quantum phenomena, the two forms are not equivalent. Suppose that instead of the quantum coin the observer uses a classical coin. Then it is \emph{in principle} possible to calculate the result of the next toss given enough initial information. Thus, in this case, the existence of the outcome of the forthcoming toss is not ruled out by the weak positivistic postulate\footnote{One might object that the required information cannot be known prior to the toss. However, we can imagine that the coin is completely shielded from the environment so that what happens in the box at time $t$ can entirely be deduced from what happens at time $t'<t$.}.

\begin{figure}[hbtp]
\begin{center}
\includegraphics{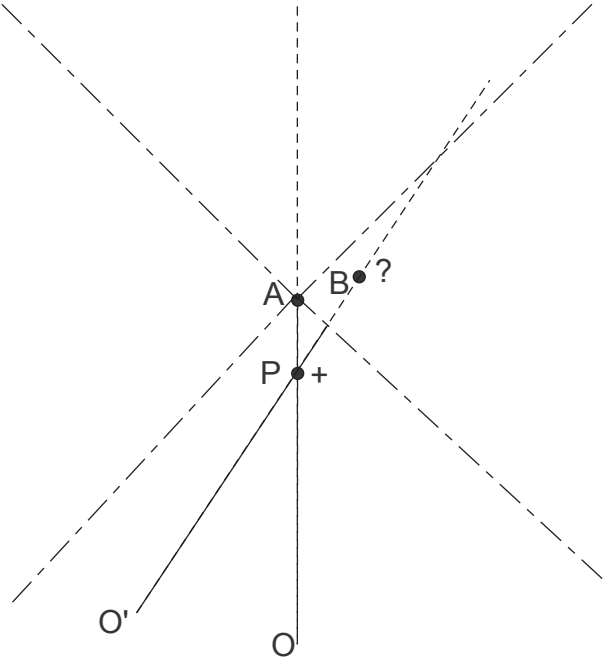}
\end{center}
\caption{The event $B$ is not real for ${\cal O}$ when ${\cal O}$ is at $A$.}\label{pileface3}
\end{figure}

We have already seen that the weak positivistic postulate and the existence of a quantum coin imply that the interior of the forward cone of an event $A$ does not exist for ${\cal O}$ at $A$. Let us generalize this result. To this aim, consider another observer, say ${\cal O}'$ who is also equipped with a quantum coin. Suppose that ${\cal O}$ and ${\cal O}'$ meet at $P$, which is before $A$ on the world-line of ${\cal O}$ (see figure \ref{pileface3}).

When ${\cal O}'$ uses the coin, this defines an event $B$. Varying the position of $P$ with respect to $A$, the speed of ${\cal O}'$ with respect to ${\cal O}$, and the delay $\Delta\tau$ before the toss, we easily see that $B$ could be anywhere in spacetime. Thus, when ${\cal O}$ is at $A$, $B$ is either inside the full backward cone of $A$, in which case the outcome of the toss can be determined by ${\cal O}$, or it is outside, in which case there is no way $A$ can determine the result. Therefore, nothing that is outside the full backward cone of $A$ can be real for ${\cal O}$ at $A$\footnote{This conclusion might seem rough. Indeed, we only proved that the state of the coin at $B$ is unreal for ${\cal O}$ at $A$, but there may be other things taking place at this point. However, the outcome of the toss can be coupled to a macroscopic system affecting everything in the forward cone of $B$. It means that the forward cone of $B$ is not real for ${\cal O}$ at $A$. Varying $B$ we get the same conclusion as before.}.

Thus, we see that special relativity, quantum mechanics, and the weak positivistic postulate, rule out every theory of time except OS presentism and OS possibilism. We think this is worth noticing that PE presentism/possibilism are not only incompatible with general relativity but also with quantum theory. It might have been anticipated that eternalism is not compatible with quantum mechanics, but it has to be emphasized that the trouble for this theory of time really comes from the combination of quantum mechanics with the weak positivistic postulate. We will see in the next section that eternalism can be founded on a quite different postulate, which is compatible with quantum mechanics.

We have seen that presentism and possibilism can be made compatible with special relativity only at the cost of renouncing to the notion of an observer-independent reality. Of course it means that those who do not wish to renounce to that notion are naturally led to eternalism by special relativity. On the other hand, the belief that the whole of spacetime exists obviously entails that reality is observer-independent\footnote{Recall that we do not consider here aspects of reality that might be other than spatio-temporal. Also, we do not wish to be entangled with the dispute over the existence of spacetime {\it per se} vs relationalism.}. In fact, general relativity makes clear that spacetime is a global entity which encompasses  many partial observations. Infinitely many different points of view fit together in a coherent whole, which, as such, does not depend on any particular point of view. Minkowski was the first to build such a coherent whole out of the different inertial frames of special relativity. Of course, such a global structure is, by definition, out of the reach of a strictly empirical, hence local, definition of reality. It is a mathematical object. The presentist/possibilist/positivist will say it is \emph{only} a mathematical object, but the eternalist will view it as representing reality more faithfully that any local observation. More simply, the eternalist will say that   this object \emph{is} reality (or at least a part of it).

Does this vision get ruined by the existence of random events ? Before answering, we should certainly have a clear mind about what randomness truly means. This is a vast subject, and we will only graze the surface of it. Nevertheless, let us remark that the very notion of probability has emerged from everyday experience, whereas everyday phenomena, as far as we know, are fundamentally governed by the deterministic laws of classical physics. Computing probabilities is just a matter of measuring the set of outcomes which is compatible with the information one has. Verifying a probabilistic prediction is just computing frequencies. Thus, the intuitive notions of probabilities, randomness, chance, and the like, may be said to have been explained by probability theory and statistics, which know nothing about probability, but only deal with measures and frequencies. This process is similar in many ways to the explanation of temperature in terms of molecular agitation. Now it is often claimed that quantum phenomena are purely random, or equivalently that quantum theory deals with pure probability. If the analogous claim that ``there exists pure temperature'' was made, it would certainly be viewed as quite extraordinary. Our aim here is not to advocate for some theory of hidden variables, for which we think that there is little hope. We want to examine what is really meant by ``purely random''. We see two ways of giving a meaning to these words (although we cannot help but thinking that only one of these two ways really makes sense !).

The first is that there are events that may or may not happen, according to some probabilistic rules. Consider for instance a quantum coin, like the one we used in the previous section. It may comes heads or tails, with equal odds. Once it is tossed, a possibility will become a reality, and the other will not. Only one of the two possible events really exists. It is apparent that this view shares many common points with presentism/possibilism: it is easy to visualize but hard to formalize, it is deeply rooted in common sense and intuition\footnote{Jokingly, one could say that a quantum coin is like a real coin, except that a real coin is not a real coin !}, and one cannot escape using terms (may, become, \ldots) that are no less obscure that what one seeks to define. This view is at the heart of the Copenhaguen interpretation of quantum mechanics.

The second way to understand ``pure probability'' in quantum theory, is to say that they are not a crude model of a more detailed reality, but that, as far as we know, they \emph{are} real. That is to say, the mathematical object which is called a probability exists in nature. It also means that its source set, usually called ``the universe'', also exists. In it, all outcomes equally exist. Consider again the quantum coin. In this view, both events ``it comes heads'' and ``it comes tails'' exists. And both are observed, but each by a different observer. In fact, the observers differ only by the outcome they observe. This view is of course the many-worlds interpretation of quantum mechanics.

Many-worlds interpretation is the obvious lifeboat for eternalism. In this view randomness is  purely subjective, as is the passage of time. From an ``outside'' point of view, nothing is random, and in fact, nothing happens, the multiverse simply is.

\begin{figure}[hbtp]
\begin{center}
\hspace{-2cm}\parbox{4.2cm}{\includegraphics[scale=0.75]{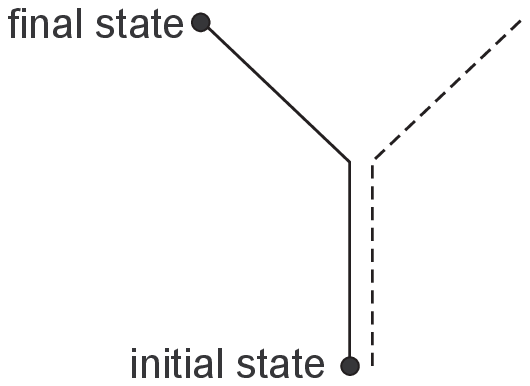}}\parbox{4.2cm}{\includegraphics[scale=0.75]{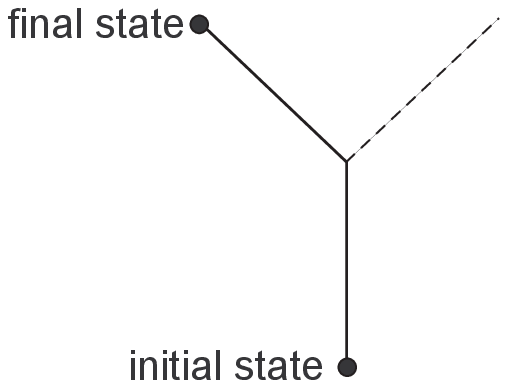}}\parbox{4.2cm}{\includegraphics[scale=0.75]{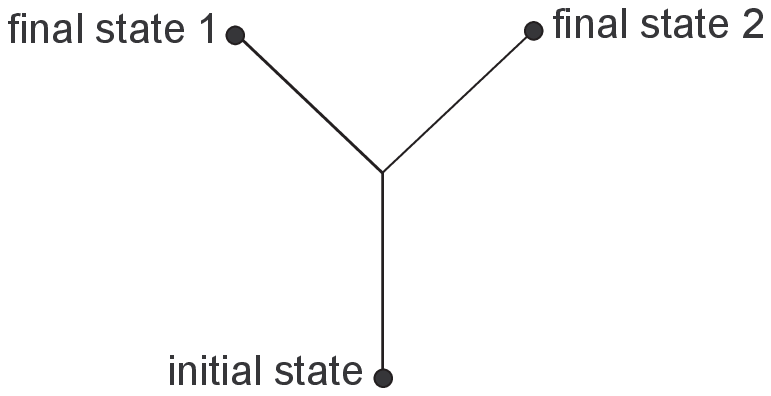}}
\end{center}
\caption{Three views on quantum mechanics: hidden variables (no true randomness), Copenhaguen interpretation (objective randomness, one world), Everett's theory (subjective randomness, many-worlds).}
\end{figure}

All this is strikingly reminiscent of the previous discussion about the existence of spacetime. Indeed, in Everett's interpretation all vectors in the Hilbert space exist alike. Reversing the chronology, one could say that Minkowski spacetime is the Everett interpretation of special relativity ! As we already noticed, guided by the postulate that there must exist an observer-independent reality, the eternalist is led to grant existence to a ``big whole'', which comprises what the possibilist/presentist would view as an infinite number of different realities. That this is consistent is guaranteed by mathematics. Indeed, this ``big whole'' is nothing more than a mathematical structure.


\section{Conclusion}

I hope that we have been able to show clearly enough that the eternalist on the one hand, and the presentist/possibilist on the other hand, are engaged in two divergent paths. For the presentist/possibilist, the passage of time is objective, and reality is subjective, whereas for the eternalist this is the other way around. Moreover, presentist/possibilist view of reality becomes more and more ontologically sparse, and less and less unified. Here again, the eternalist stance moves in the opposite direction. 

Indeed, as we have seen, the presentist/possibilist is forced by modern physics to adopt a definition of reality which depends on the immediate perceptions of the observer. Moreover, we have argued that there is no sensible notion of present in general relativity, even though the possibilist reality (present + past) seems to still make sense, at least if some conditions are physically satisfied (no CTC). This purely subjective perspective on reality is incarnated by the positivistic postulate. Moreover we have seen that the existence of an open future, which is one of the main arguments for presentism/possibilism, relies on the intuition that pure probabilities exist, and this boils down to the Copenhagen interpretation of quantum mechanics, in which observations play a special role. Thus, in order to be compatible with modern physics, these theories of time (in fact, only one of them survives if our arguments are correct) acquire some very counter-intuitive, seemingly paradoxical, elements (the same event, for instance my own birth, may be present several times, but on the other hand, what happens at the other side of the planet ``now'', as defined by the usual Poincar\'e-Einstein convention, does not exist). Even though it does not render them  inconsistent, one must admit that their closeness to common sense breaks down, so that they lose most of their  initial appeal. We think that this is the sign that these theories use  concepts which take their origin in everyday experience, and are relevant at this level, but are not well suited to think about the fundamental nature of time. The same sort of thing happens when one tries to explain quantum phenomena in terms of wave or particles.

On the other hand, the eternalist reality must accommodate a whole multiverse in order to be compatible with quantum mechanics. In fact, general relativity already shows that the existence of the totality of spacetime entails the existence of forever inaccessible regions. Thus, the eternalist manages to maintain his position at the cost of a vast enlargement of reality. Will this view someday be destroyed by a new physical theory ? We claim that it cannot happen, granted that it will always be possible to define a single mathematical object comprising all mathematical entities entering the formulation of the theory. Historically, this has always proven to be feasible (Minkowski spacetime for special relativity, the Hilbert space of states for quantum theory, nets of $C^*$-algebras for quantum field theory\footnote{The last example is certainly not as thoroughly established as the former two.}, etc.). We do not see any indication that it should be otherwise in the future. Thus, we think that there will always be an Everett-style interpretation of the theory. Of course, this is an act of faith. However, it is possible to settle that question by subsuming eternalism into mathematical realism.

In a sense, mathematical realism, the idea that mathematical objects are real, can be seen as a maximal extension of eternalism. Indeed, we have seen that eternalism, under the pressure of modern physics, must grant physical reality to more and more objects which, for a presentist or possibilist, would   exist only  mathematically. Going directly to where that road seems to be going, it appears natural to admit that all mathematical objects exist. One could reply that an eternalist need not commit to the existence of any mathematical object. For instance, a spacetime point might be said to represent a part of reality without being real by itself. In short, we could be accused of confusing reality and some representation of it. However, as far as relativity is concerned, there is nothing more to reality than a Lorentz manifold. Admittedly, this does not cover every aspect of reality, but this must be seen as a limitation of this particular physical theory rather than as a proof of any essential difference between mathematical existence and physical existence. If we someday formulate a theory which correctly predicts all experiments and explains all observations, how would we make a difference between reality and the theory ? From this, we ask: \emph{why} should we make a difference ? Should we not use some form of ``Turing test'' here ? Still, we think that there is some relevance to the point that we should be cautious not to confuse reality and its representation. Not to establish a frontier between mathematics and physics, but inside mathematics themselves. We will elaborate on this below.

Of course, some might resist the conclusion that eternalists should commit to mathematical realism. Truly, the question about the reality of past and future events is different from the question ``are mathematical objects real ?''. But both questions should be answered immediately once a definition of reality is given. It would even seem more logical to begin a discussion about the reality of past and future events by a definition of reality itself. However, any definition would probably sound arbitrary. In fact, this article may be seen as an attempt to motivate two definitions of reality, one   well-suited to possibilism/presentism, and the other to eternalism. The first definition has been already given, or nearly so in the form of the positivistic postulate. One just has to turn the implication into an equivalence: 

\begin{quote}
{\bf Presentist/possibilist definition of reality:} \emph{The reality of ${\cal O}$ at some moment is the totality of what, in principle, ${\cal O}$ can be observing/can have observed.}
\end{quote}

We note that, thanks to the words \emph{in principle}, we do not need to take into account the actual possibilities of observation (whether the propagation of light, or any other information carrier, is slowed down or even blocked by some medium). Thus, we can readily give a precise physical meaning to this definition by making use of the causal structure of spacetime \emph{and} of a time arrow. 

As for the eternalist definition, we have seen that it should possess the property of observer-independence, and argued that is should include some, and maybe all mathematical objects. We are just one step away from what seems to be a consistent definition:

\begin{quote}
{\bf Eternalist definition of reality:}\emph{ Reality is the totality of mathematical objects.}
\end{quote}

Now that we have come to this conclusion, it is time to turn the problem upside down and check consistency. Does the presentist/possibilist definition of reality  implies presentism/possibilism ? Obviously, the presentist/possibilist definition does imply, tautologically, that only the present/past, in its OS form, exists.  However, we have seen that general relativity implies that such a notion of the present conflicts with the idea that time passes, which is also a component of presentism. Thus we see that, in the context of general relativity, presentism is not consistent. We have also seen that the possibilist definition does make sense as long as closed timelike curves are forbidden. Thus, the consistency of the definition is not guaranteed: it depends on the latest physical theory\footnote{It might even serve as a guiding principle for future theories.}. Thus, in a way, it is refutable.  

It should be obvious that the eternalist definition implies eternalism. Moreover, not only is this definition compatible with modern physics, but it shall always remain so, provided physics will always be mathematically formalized. Some will certainly wince at a definition of reality which is impervious to future discoveries. However, it is not intended to be a scientific theory. Anyway, it has many advantages from a philosophical point of view. First, it deals with the issue of time without any reference to time. Moreover, the eternalist definition gives answers to other problems. Since mathematics know nothing about randomness, free will, or becoming, the questions about the objective reality of these notions is answered in the negative. Another interesting application of this definition is that it implies  Everett's interpretation of quantum mechanics. Finally, it solves the issue raised by Wigner about the ``unreasonable efficiency of mathematics''.

Some might be worried about that definition dealing so easily with ancient problems. Are we not ``trivializing'' everything ? I do not think so, since the question ``what is a mathematical object ?'' has no obvious answer\footnote{Note that neither does the question ``what is an observation ?''.}. This is the place to be cautious about the distinction between reality and the representation of it, because mathematical objects, whatever it may mean,  have a lot of equivalent representations, which may seem to be of very different essences. To give just one example, a point in a compact topological space can be viewed as a character on a commutative $C^*$-algebra. In fact, playing with the many different facets of the same mathematical reality is a practice which is both common and fruitful for mathematicians. Although this is not the place to discuss such matters, most mathematicians would certainly agree that their work is to uncover the mathematical reality which is hidden behind all these different guises.

Thus, to those who might object that reality should not be defined {\it a priori}, that it is the role of \emph{physics} to define it, we would reply that the eternalist definition differs on just one point from their view: it puts the burden on mathematics rather than physics.

\bibliographystyle{alpha}
\bibliography{bibliotime}

\begin{thebibliography}{McT08}

\bibitem[Bes11]{bes}
F.~Besnard.
\newblock Simultaneity in {M}inkowski spacetime: from uniqueness to
  arbitrariness.
\newblock \url{http://arxiv.org/abs/1104.1263}, 2011.

\bibitem[Big66]{bigelow}
J.~Bigelow.
\newblock Presentism and properties.
\newblock In James~E. Tomberlin, editor, {\em Philosophical Perspectives 10:
  Metaphysics}. 1966.

\bibitem[BY06]{benyami}
H.~Ben-Yami.
\newblock Causality and temporal order in special relativity.
\newblock {\em Brit. J. Phil. Sci.}, 57:459--479, 2006.

\bibitem[Car63]{carnap}
R.~Carnap.
\newblock {\em The Philosophy of Rudolf Carnap}, chapter Intellectual
  autobiography.
\newblock LaSalle: Open Court, 1963.

\bibitem[Cra00]{craig}
W.~Craig.
\newblock {\em The Tenseless Theory of Time: A Critical Examination}.
\newblock Dordrecht: Kluwer Academic Publishers, 2000.

\bibitem[Ein01]{eins}
A.~Einstein.
\newblock {\em Relativity: The Special and General Theory}.
\newblock Routledge, 2001.

\bibitem[Giu01]{giul}
D.~Giulini.
\newblock Uniqueness of simultaneity.
\newblock {\em British Journal for the Philosophy of Science}, 52:651--670,
  2001.

\bibitem[GS79]{godsmith}
W.~Godfrey-Smith.
\newblock Special relativity and the present.
\newblock {\em Philosophical Studies}, 36:233--244, 1979.

\bibitem[Haw92]{hawking}
S.~Hawking.
\newblock The chronology protection conjecture.
\newblock {\em Phys. Rev. D}, 46:603--611, 1992.

\bibitem[H.W09]{weyl}
H.Weyl.
\newblock {\em Philosophy of Mathematics and Natural Science}.
\newblock Princeton University Press, revised edition, 2009.

\bibitem[Jan10]{janis}
A.~Janis.
\newblock {\em Conventionality of simultaneity}.
\newblock The Stanford Encyclopedia of Philosophy (Fall 2010 Edition), 2010.

\bibitem[JB81]{beem}
P.~Ehrlich J.~Beem.
\newblock {\em Global Lorentzian Geometry}.
\newblock Marcel Dekker: New York, 1981.

\bibitem[JF90]{novikov}
I.~Novikov F. Echeverria G. Klinkhammer K. Thorne U.~Yurtsever J.~Friedman,
  M.~Morris.
\newblock Cauchy problem in spacetimes with closed timelike curves.
\newblock {\em Physical Review D}, 42:1915--1930, 1990.

\bibitem[Mal77]{malam}
D.~Malament.
\newblock Causal theories of time and the conventionality of simultaneity.
\newblock {\em No\^us}, 11:293--300, 1977.

\bibitem[McT08]{mactag}
J.~M.~E. McTaggart.
\newblock The unreality of time.
\newblock {\em Mind, New Series}, 68:457--484, 1908.

\bibitem[Min52]{mink}
H.~Minkowski.
\newblock Space and time.
\newblock In {\em The Principle of Relativity: A Collection of Original Memoirs
  on the Special and General Theory of Relativity}, chapter Space and Time.
  Dover, New York, 1952.

\bibitem[Pen89]{penrose}
R.~Penrose.
\newblock {\em The Emperor's New Mind: ``Concerning Computers, Minds, and Laws
  of Physics"}.
\newblock New York and Oxford: Oxford University Press, 1989.

\bibitem[Pet06]{petkov}
V.~Petkov.
\newblock Is there an alternative to the block universe view ?
\newblock In M.~Redei D.~Dieks, editor, {\em The Ontology of Spacetime}.
  Elsevier, Amsterdam, 2006.

\bibitem[Pri70]{prior}
A.~Prior.
\newblock The notion of the present.
\newblock {\em Studium Generale}, 23:245--248, 1970.

\bibitem[Put67]{putnam}
H.~Putnam.
\newblock Time and physical geometry.
\newblock {\em Journal of Philosophy}, 64:240--247, 1967.

\bibitem[Rei69]{reich}
H.~Reichenbach.
\newblock {\em Axiomatization of the theory of relativity}.
\newblock Berkeley University Press, 1969.

\bibitem[Rie66]{riet}
C.~W. Rietdijk.
\newblock A rigorous proof of determinism derived from the special theory of
  relativity.
\newblock {\em Philosophy of Science}, 33:341--344, 1966.

\bibitem[RS77]{sachs}
H.~Wu R.K.~Sachs.
\newblock {\em Relativity for Mathematicians}.
\newblock Springer Verlag: Berlin, Heidleberg, New York, 1977.

\bibitem[Sau02]{saunders}
S.~Saunders.
\newblock How relativity contradicts presentism.
\newblock In C.~Callender, editor, {\em Time, Reality and Experience}.
  Cambridge University Press, 2002.

\bibitem[Sav00]{savitt2}
S.~Savitt.
\newblock There's no time like the present (in {M}inkowski spacetime).
\newblock {\em Philosophy of Science}, 67:S563--S574, 2000.

\bibitem[Sav06]{savitt}
S.~Savitt.
\newblock Being and becoming in modern physics.
\newblock In {\em The Stanford Encyclopedia of Philosophy}. 2006.

\bibitem[Sch08]{schechter}
P.~Schechter.
\newblock The {H}ubble constant from gravitational lens time delays.
\newblock In {\em Proceedings of IAU Symposium No. 225: The Impact of
  Gravitational Lensing on Cosmology}, 2008.

\bibitem[SS99]{sarkstach}
J.~Stachel S.~Sarkar.
\newblock Did {M}alament prove the nonconventionality of simultaneity in the
  special theory of relativity ?
\newblock {\em Philosophy of Science}, 66:208--220, 1999.

\bibitem[Ste68]{stein}
H.~Stein.
\newblock On {E}instein-{M}inkowski space-time.
\newblock {\em The Journal of Philosophy}, 65:5--23, 1968.

\bibitem[YB03]{bala}
M.~Janssen Y.~Balashov.
\newblock Presentism and relativity.
\newblock {\em British Journal for the Philosophy of Science}, 54(2):327--346,
  2003.

\end{thebibliography}

\end{document}